\documentclass[fleqn,twoside]{article}

\usepackage{espcrc2}

\usepackage{graphicx}

\def\ARAA{{\it Annual Rev. of Astron. \& Astrophys.} }

\def\ApJ{{\it Astrophys. J.} }
\def\ApJL{{\it Astrophys. J. Letters} }

\def\ApP{{\it Astropart. Phys.} }
\def\AA{{\it Astron. \& Astroph.} }
\def\AAR{{\it Astron. \& Astroph. Rev.} }
\def\AAL{{\it Astron. \& Astroph. Letters} }

\def\JHEP{{\it Journal of High Energy Physics} }
\def\JPhG{{\it Journ. of Physics} {\bf G} }

\def\PRD{{\it Phys. Rev.} {\bf D} }
\def\PRL{{\it Phys. Rev. Letters} }

\def\MNRAS{{\it Month. Not. Roy. Astr. Soc.} }
\def\ZA{{\it Zeitschr. f{\"u}r Astrophys.} }

\def\etal{{\it et al.}}

\newcommand{\AmS}{{\protect\the\textfont2

  A\kern-.1667em\lower.5ex\hbox{M}\kern-.125emS}}

\title{ULTRA HIGH ENERGY COSMIC RAY SOURCES \& EXPERIMENTAL RESULTS}
\author{Peter L. Biermann$^{1,2}$, Gustavo Medina Tanco$^3$
\\[10mm]
$^1$Max-Planck Institute for Radioastronomy, Bonn, Germany \\
$^2$Department for Physics and Astronomy,\\ University of
Bonn, Germany\\
$^3$Instituto Astronomico e Geofisico\\
Universidade de Sao Paulo, Brasil\\
plbiermann@mpifr-bonn.mpg.de,gustavo@astro.iag.usp.br\\[10mm]
}

\begin{document}

\begin{abstract}

Here we discuss the latest developments in the debate, where the
ultrahigh energy cosmic ray particles come from.  In this brief review,
we emphasize the predictions that necessarily follow from the various
concepts proposed.  We discuss both sources and propagation, and spend
some space on the by now most conservative model, giant radio galaxies
and their hot spots and jets, because it allows more steps in the
reasoning to be checked, and is more tightly constrained than any other
model.  It has survived about forty years of debate already.  We
summarize the task ahead of us at the end, including the development of
new technologies to observe ultra high energy cosmic rays, e.g., with the
radio emission of air showers, which may lead to an alternative to the
fluorescence detectors with their small duty cycle.  Due to HiRes, AUGER,
and the plans for EUSO, the future is bright, and the ultra high energy
cosmic ray particles may yet allow us to explore new physics.

\vspace{1pc}

\end{abstract}


\maketitle

\section{Introduction}

Over the last few years controversies have arisen both about the
experimental results, and also about the theoretical interpretations.  We
are going to review the situation, but will do so in very limited space.

\subsection{The cosmic ray spectrum}

The cosmic ray spectrum in nuclei shows the following basic properties:

\begin{itemize}

\item{}  approx. $E^{-2.7}$ until the knee

\item{}  a bend downwards at around $3 \, 10^{15}$ eV

\item{}  approx. $E^{-3.1}$ beyond the knee, with a slight downward dip
from $3 \, 10^{17}$ eV

\item{}  a transition near $3 \, 10^{18}$ eV, and then approx.
$E^{-2.7}$ again

\item{} uncertainty beyond $5 \, 10^{19}$ eV, either mild cutoff (HIRES)
or a continuation (AGASA)

\end{itemize}

A main reference for a collection of data until 1997 is in
\cite{LB-CR}. General references are in
\cite{GS63,Gaisser90,Venyabook,B97}.  Classical reviews are,
e.g., in \cite{Hillas84}, while some recent reviews are
\cite{Bhatta99,NW2000,BS01,LS01,BiSi01,Ha02}.

Generally we emphasize very recent work, and suggest the reader consult
recent reviews for older references.

\subsection{Interactions in the microwave background}

The ultra high energy cosmic ray particles interact with the microwave
background, and their losses lead to an ubiquitous cutoff, usually
referred to as Greisen-Zatsepin-Kuzmin cutoff, after its discoverers,
under some simplifying assumption about the nature of these particles
(protons), the nature of intergalactic propagation, as well as source
distribution.

\begin{itemize}

\item{}  pair and pion production

\item{}  limits effective distance from which a high energy proton (or
neutron, or nucleus) may come

\item{}  discovered by K. Greisen, G. Zatsepin and V. Kuzmin

\item{}  GZK distance about 50 Mpc, a travel distance

\item{}  what sources?

\end{itemize}

The key references are \cite{Greisen66,ZK66,NW2000}.  A recent
generalization to nuclei is in \cite{ER98a,ER98b,EMR99,BILS02}.   A
popular account is in \cite{Clay98}.

\section{Galactic Cosmic Rays}

\subsection{Transition from Galactic Cosmic Rays?}

The newest data suggest and confirm that the transition from Galactic
cosmic rays to an extragalactic population appears to be near $3 \,
10^{18}$ eV.

\begin{itemize}

\item{}  Near $3 \, 10^{18}$ eV transition

\item{}  below heavy to medium nuclei, above light nuclei

\item{}  spectral dip there seen by Fly's Eye

\item{}  now confirmed by AGASA, HiRes, Yakutsk,...

\end{itemize}

\subsection{Origin of Galactic Cosmic Rays}

A quantitative theory for Galactic cosmic rays was proposed in 1993,
\cite{CRI,CRII,CRIII,CRIV,CalgaryCR,TucsonCR,CR9}, with an earlier step
\cite{HJV+PLB88}.  Quantitative tests are now again possible with the
KASKADE data, as well as the latest HiRes, Akeno, Yakutsk, and Haverah
Park data.  Detailed simulations have been carried out at several levels
of sophistication by A. Vasile and S. Ter-Antonyan, and they show
consistently, that the older proposals do give a suitable interpretation
of the new data at all energies for Galactic cosmic rays.  At the level
of today's knowledge the older proposal thus does find confirmation in
the new data.  The suggestion as how to explain the knee,
\cite{Peters59,Peters61}, based on a slightly diminished efficiency of
curvature drift in the supernova shock racing through a magnetic wind of
a massive star, strongly points to a specific mechanism for the
supernova, the mechanism proposed by G. Bisnovatyi-Kogan in 1970, based
on earlier ideas by N. Kardashev from 1964,
\cite{Ka64,Genn71}.  In these ideas the core of the star collapses to a
small disk, held by the angular momentum barrier, and then magnetic
fields transport energy and angular momentum to the envelope, exploding
the star, and allowing final collapse of the core.  Such a mechanism
naturally gives the correct energies -
$10^{52}$ erg is required for the explosion, and seen in Supernova 1998bw,
\cite{SN1998bw-1,SN1998bw-2}.  If accepted, then this in turn suggests
that all very massive stars come to a common final end at the point of
supernova explosion, and there seems a possibility that they could become
a new and very bright standard candle in cosmology - provided we find a
correction for the extreme asphericity expected from the Bisnovatyi-Kogan
mechanism.  In this context it is also possible perhaps to understand the
gamma ray emission of the Galaxy,
\cite{Hu97} as well as the B/C ratio energy dependence, \cite{Pt99,CR9};
as the general activity in the Galaxy in almost everything connected to
young and massive stars is centered in the Galactic center region, one
might expect that that region may serve as a paradigm for bot stellar and
perhaps even non-stellar types of activity,
\cite{Mezger96,MF2001}.  Other concepts are explored in
\cite{CER02,CRE02,CMR02}.

\begin{itemize}

\item{}  consistent with (KASKADE, HiRes, Akeno,...):

\item{}  Red Supergiant (RSG) stars and Wolf Rayet (WR) stars
explode, and Kolmogorov spectrum in Inter-Stellar Medium (ISM)

\item{}  knee from transition in efficiency of curvature drift
acceleration

\item{}  final cutoff due to magnetic field in wind, Larmor limit

\item{}  consequences for Supernova mechanism, and possibly bright
standard candle

\end{itemize}

Details will be published in further papers.

\section{Homogeneous source distribution?}

The GZK cutoff is expected under several assumptions, one of which is a
homogeneous source distribution.  There are no astronomical sources
known, which are truly distributed in a homogeneous fashion.

\begin{itemize}

\item{}  The very high energies, above $3 \, 10^{18}$ eV:

\item{}  assume zero effect of magnetic field

\item{}  assume protons (or nuclei)

\item{}  assume an initial energy far beyond anything observed

\item{}  obtain a strong downturn feature: The GZK-cutoff expected near
$5 \, 10^{19}$ eV

\end{itemize}

The latest data from HiRes and AGASA do not confirm each other on whether
the expected cutoff has been seen at all, and also do not clearly
show whether the data are compatible with the simple cutoff,
\cite{HiRes02,Takeda02,BaWa02}.  The very existence of events beyond
$10^{20}$ eV suggests that the very simple model cannot work, and one of
the key points was the assumed homogeneous source distribution.  Already
a source distribution more closely consistent with bright and early
Hubble type galaxies suggests a clearly higher expected flux near
$10^{20}$ eV, e.g., \cite{BBO01}.

\section{Sky distribution?}

For many years a controversy has raged on whether specific events do
correlate with known sources; one of the earliest suggestions by Ginzburg
was that the nearest powerful radio galaxy, M87, could be a good
candidate - it still is. However, a direct positional coincidence of an
event with a candidate source, or better, with an entire class of
sources, has been argued repeatedly, \cite{FB98}.  Then the claim by
AGASA, that there are double and triple events, does allow, again under
very simplifying assumptions, to obtain statistics on possible sources.
Recent investigations of such correlations are
\cite{Gl01a,Gl02a,Gl02b,Gl02d,Gl02e}
and \cite{Dubovsky00,Tinyakov01a,Tinyakov01b,Tinyakov02a,Gorbunov02}.

\begin{itemize}

\item{}  Assume no magnetic field

\item{}  then sky distribution: many sources

\item{}  detection of doubles and triplets: real sources

\item{}  are there sources there?

\item{}  with enough power?  $L_{CR} \; = \; L_{el.magn.}$ at most

\item{}  Using the deep radio surveys, there is always such a source

\end{itemize}

Under the proposal, that radio galaxies such as M87, a source, for which
the observed optical spectrum suggests the presence of protons of
$10^{21}$ eV \cite{BS87}, required to explain the very common cutoff in
the optical synchrotron emission near $3 \, 10^{14}$ Hz, are good
candidate sources, all radio galaxies, and a fortiori, all BL Lac type
sources are candidates, see, e.g., \cite{PDR02}.  And, under the standard
unification picture of active galactic nuclei, in the same vein, all flat
spectrum radio sources are good candidates, under this by now most
conservative model to explain ultra high energy cosmic rays.  It is not
surprising that such claims have been tested, in various papers such as
\cite{FB98}, and now, in a new series of papers by P. Tinyakov \&
I. Tkachev,
\cite{Dubovsky00,Tinyakov01a,Tinyakov01b,Tinyakov02a,Gorbunov02}, and
others, \cite{STAR01}.  All these papers find tantalizing hints of a
possible correlation between sources in this class, and observed cosmic
ray events; however, the interpretation requires uncharged particles with
a penetrating power through the bath of the microwave background far in
excess of what protons can do.  Such particles may well exist, but there
is no self-consistent theory for them at this time.  Also the statistics
of source correlations have not been confirmed to everybody's
satisfaction.  On the other hand, any highly significant correlation of
even a subset of all observed events with a known class of sources
is interesting, and a clear clue to any real source, as well as to the
nature of particle.

The Yakutsk data and an extensive discussion of the array and an
interpretation of its data have been published in a series of papers,
\cite{Gl00,Gl01a,Gl01b,Gl02a,Gl02b,Gl02c,Gl02d,Gl02e,Mi02}, but the
conclusion is not ultimately clear; we need to understand the experiment
better in order to compare its important results with those of other
experiments.

\section{Magnetic Fields}

Magnetic fields in the universe are known to exist, now for over fifty
years, \cite{Kronberg94,Kulsrud99}, and they seem to be just about
everywhere, \cite{RKB98,Kronberg99,CKB2001}.

\begin{itemize}

\item{}  Magnetic fields have been detected

\item{}  they are strong

\item{}  they are highly inhomogeneous

\item{}  they compete with other forms of energy

\item{}  only in cosmic clusters, filaments and sheets,

\item{}  but not in the voids, and not overall

\item{}  5-10 microGauss in clusters - at least

\item{}  0.01 to 1 microGauss in filaments and sheets

\item{}  bending of paths of energetic particles

\end{itemize}

Therefore, magnetic fields cannot be ignored in the propagation of
energetic particles, and even at the highest energies observed the
cosmic magnetic fields have a noticeable effect.  In fact, magnetic
fields correlate well with the cosmological galaxy distribution, and so
are relatively strong in the sheets and filaments of the galaxy
distribution, and so also in the sheet connecting us to the Virgo
cluster.  For any particles emitted by some source in the general region
of the Virgo cluster - such as M87 - the arrival directions outside the
Galaxy are expected to cluster along the supergalactic plane then, and so
represent an elongated band in the sky - before they pass into the
magnetic field surrounding our Galaxy.  Recent work is in
\cite{HMR99,HMR00a,HMR00b,Si01b,HMR02a,HMR02b,IS02,YNTS02,BILS02}.

\subsection{Galactic Magnetic Field}

Right after the discovery of the Solar wind in 1950, \cite{LB51}, the
suggestion came up that our Galaxy may also have a wind.  If our Galaxy
has a wind, it is likely that it is driven by cosmic rays, because they
need to deposit their energy lost in adiabatic losses into the surrounding
magnetic medium, and that easily drives a wind, e.g.,
\cite{Galwind91,Galwind93,Galwind97,Galwind99}.  The driving of a
magnetic wind by cosmic rays has some similarity to the possible
influence of radiation driving a magnetic wind, \cite{SB97}. Such a wind
implies that the magnetic field topology right from the start just above
the hot thick disk is in the asymptotic E. Parker from of 1958,
\cite{Parker58}, because otherwise the wind would take so much angular
momentum out from the gas in the disk, that the Galaxy would long have
stopped all its star formation - all the gas would have been deposited in
the central region, and to some small measure in the central black hole.
This implies the topology of the magnetic field in the wind:

\begin{itemize}

\item{}  If Galaxy has wind

\item{}  driven by cosmic rays, and supernova shocks

\item{}  then magnetic field topology as in Solar Wind

\item{}  Archimedean Spiral, i.e. $B_{\phi} \; \sim \; sin \theta /r$

\item{}  turbulence spectrum $k^{-2}$ in wavenumber for adopted isotropy
in wave number phase space, derived from the saw-tooth pattern of shock
driving

\item{}  if so, then power of wind should scale with 60 micron
luminosity, since Cosmic Ray activity scales with star formation, and
star formation is visible at 60 micron

\end{itemize}

It can be shown, that all such winds can easily provide the universe
with all the observed magnetic fields, \cite{RKB98,Kronberg99,AES02}.  If
we could only understand where galaxies really get their magnetic fields
from, we would be much better shape.  Magnetic fields will be
discussed elsewhere.

\section{Predictions:  Double and triple events}

In this paper, we wish to emphasize the predictions, that arise from the
various source and propagation arguments.

In such a picture of a magnetic wind of the Galaxy, there is one obvious
explanation for the doubles and triples observed by AGASA:  magnetic
lensing, or in other words, the caustics given by the propagation of
charged particles through a magnetic wind of the Galaxy; these caustics
surely overlap for different particle energies.  Doubles and triples can
occur when there are two possible paths for particles at very different
energies, leading to the same direction observed at Earth, but coming
from different sources in the sky, presumably a band in the sky in
arrival direction as seen outside the Galaxy.  For this argument it does
not actually matter, where exactly the particles are coming from.  Recent
work on this is in \cite{HMR00b,HMR02a,SEAS02}.

\begin{itemize}

\item{}  Magnetic caustics:  $E_2/E_1$ either near 1 or from about 2-3

\item{}  Then galactic wind required, driven by cosmic rays

\item{}  and permeated by a relatively strong magnetic field

\item{}  Sources:  Reasonably sampled continuous spectrum - implies
neutral particles

\item{}  Requires strong source

\end{itemize}

\begin{figure}
\centering\rotatebox{0}{\resizebox{8cm}{!}%
{\includegraphics{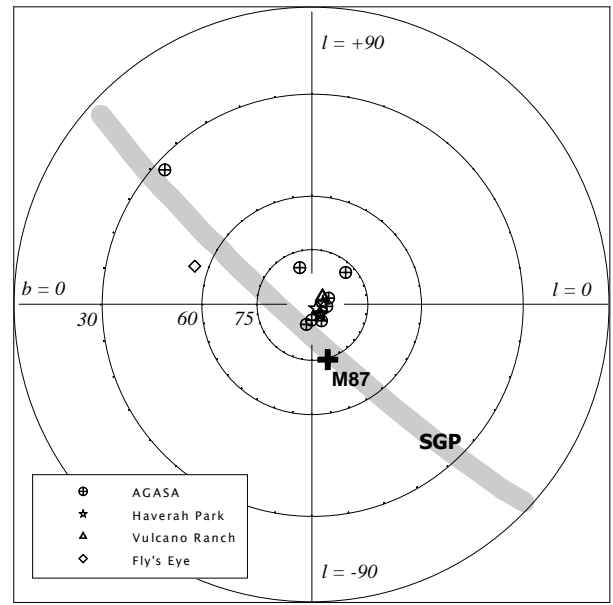}}} \caption{In this graph we
show the arrival directions of observed cosmic rays calculated
back to outside the assumed Galactic wind model with a Parker like
magnetic field topology }
\end{figure}

\begin{figure}
\centering\rotatebox{90}{\resizebox{10cm}{!}%
{\includegraphics{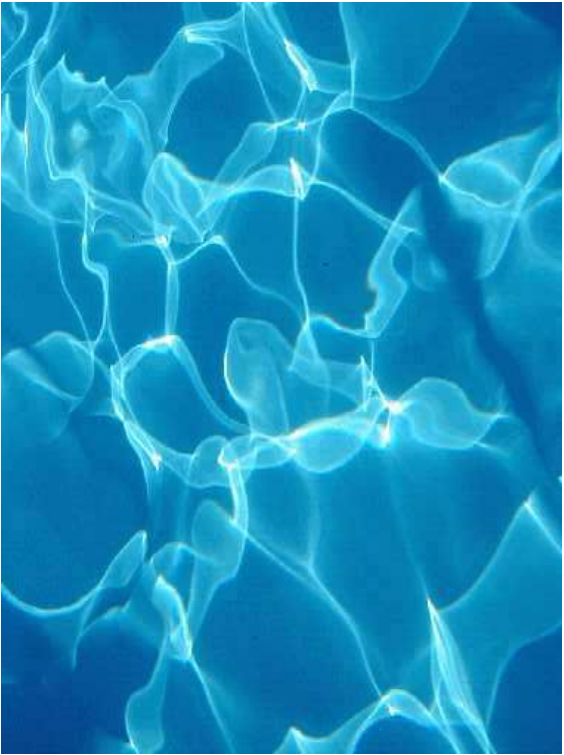}}} \caption{In this graph
magnetic lensing is shown in a simulation based on sunlight
shining on a swimming pool, and then considering the pattern of
light at the bottom of the pool.  This clearly shows the caustics
in the imaging, and the peaks of light, corresponding possibly to
the doubles and triples of events observed. Here we need to
imagine a similar pattern, derived basically by scaling; then the
superposition of such patterns at various scalings to represent
different particle energies allows overlaps, which could be
interpreted as those sites where we have mutliple events at very
different particle energies.  This graph is due to E. Roulet}
\end{figure}

\begin{figure}
\centering\rotatebox{90}{\resizebox{13cm}{!}%
{\includegraphics{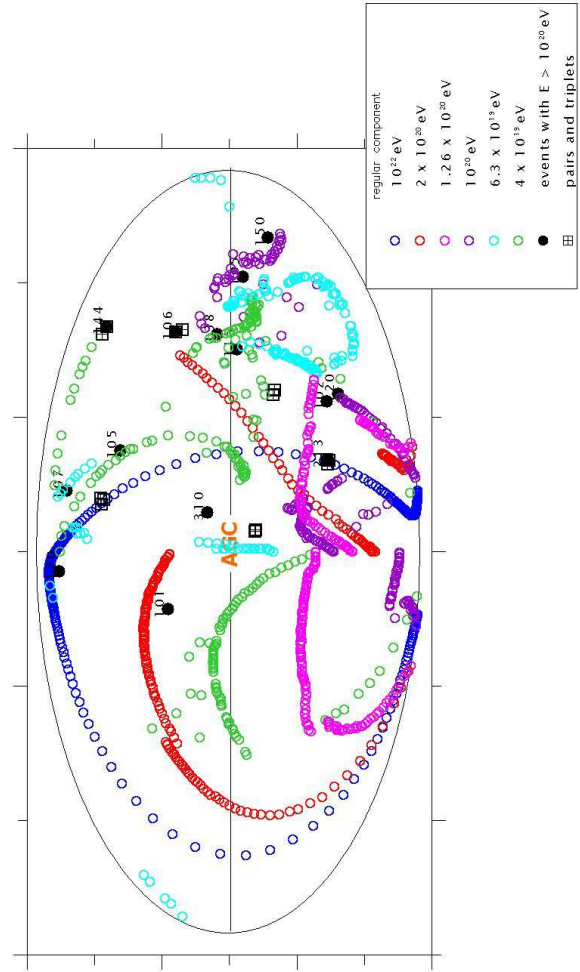}}} \caption{In this graph we show
the arrival directions of cosmic rays at various energies, coming
from the supergalactic plane in a long band, and then down to us
through a Galactic wind modeled after Parker again.  At the
highest energy, very little bending is shown - a factor of 30
above the maximum energy observed.  As then the energy of the
particle used comes down, the bending becomes first noticeable,
and then strong, leading to overlaps between different distorted
bands.  These overlaps then can directly be interpreted as
locations on the sky, where multiple events with very different
energies are likely to occur.  In fact, the observed multiple
events are already well approximated in this simple scheme.}
\end{figure}

\section{Radio galaxies, Microquasars, Gamma Ray Bursts}

Diffusive shock acceleration is one commonly accepted now classical
mechanism to accelerate charged particles, based on the original ideas
of E. Fermi (1949, 1954)
\cite{Fermi49,Fermi54,Bell78,Lagage+C83,Drury83,Jokipii87,Venyabook}.  At
a shock the magnetic irregularities on both sides of the shock play a
relativistic tennis game with the charged particles, and, just as in real
life occasionally drop a ``ball", i.e. a particle, which then escapes
downstream from the shock.  In is most simple form this process produces
an $E^{-2}$ spectrum.

\begin{itemize}

\item{}  (weakly) relativistic shock wave acceleration

\item{}  $10^{21}$ eV proton energy in the jet of M87 required by optical
observations, \cite{BS87}

\item{}  as in many Active Galactic Nuclei (Jets, Hot Spots, or compact
sources)

\end{itemize}

The maximum particle energy scales with the root of the power in the jet,
\cite{Lovelace76}, and so low power sources such as Microquasars cannot
yield really high energies; this can be compensated by relativistic
boosting, as in Gamma Ray Bursts,
\cite{Piran99}, but then the observed fluence distribution does not give a
sufficient number of such sources in cosmological proximity to explain
either the bulk of Galactic cosmic rays nor extragalactic cosmic rays,
\cite{GRB1,GRB2,GRB3}.  A different point of view is in
\cite{Wa02,DGWL02}.  A variant on this topic is to consider magnetars,
\cite{Ar02}, or pulsars, \cite{BePr02}.  Such stellar sources
may also contribute here, but with the particles coming from other
galaxies; various options are discussed, e.g., by \cite{SGM02}.
Radiogalaxies are considered again in a simple model in
\cite{BGG02}.

\subsection{Predictions:  Microquasars and Gamma Ray Bursts}

Even though it appears unlikely hat microquasars and Gamma Ray Bursts do
provide the bulk of the observed cosmic rays at any energy, they are
clearly sources of energetic particles, and as such should contribute at
some level; as they almost certainly have a different spectrum, and a
different source distribution compared to, e.g., large early Hubble type
galaxies, their flux contribution should be examined.  It might be
visible in some special feature in cosmic rays, not yet ascertained, in
the gamma ray spectrum from galaxies, or in neutrinos - the best way to
connect to specific sources. We have a fairly good handle on the physics
of microquasars from the work by S. Markoff and G. Romero - we need to
build on this understanding, e.g., \cite{MFF01,RKBM02}.

\begin{itemize}
\item{}  also sources of energetic particles

\item{}  if pointed at us, very interesting

\item{}  unlikely to contribute at the highest energies

\item{}  doubtful to dominate at lowest energy

\item{}  could be checked in neutrinos
\end{itemize}

Recent work on the neutrino aspect is in \cite{AH01,AH02,De02}.

\section{Predictions:  M87}

To be specific, the proposal has been made \cite{BS87}, based on optical
spectra, that M87 not only requires protons in the source at $10^{21}$
eV, but also, that M87 may be the dominant source to explain the observed
ultra high energy cosmic rays.  Recent work on M87 is in \cite{PDR02}.
This leads to some clear predictions:

\begin{itemize}

\item{}  M87: all events should point back to Supergalactic Plane Sheet -
to within the noise produced in the propagation through the $k^{-2}$
turbulence spectrum in magnetic irregularities

\item{}  arrival directions seen outside our Galaxy elongated along
sheet

\item{}  transition to light elements (H, He) near and above $3 \,
10^{18}$ eV

\item{}  and such all the way to maximum energy

\end{itemize}

Clearly, M87 cannot be blamed for everything, but at the maximum energy
there may only be one source remaining among all the contributors, and
a suggestion is that this last remaining single source is M87.  Weaker
sources perhaps are other lower power radio galaxies such as Cen A, see
\cite{ILS02}, which are expected to contribute flux to about $5 \,
10^{19}$ eV, from the relation between jet power, magnetic fields and
maximum particle energy already quoted above.  Of course, it has become
obvious from the latest simulations that radio galaxies are highly time
dependent in all their manifestations, and so we need to keep the option
open, that such radio galaxies were more powerful at the time of particle
production; their past behaviour can probably be checked with X-ray data.

\subsection{Predictions:  Galactic wind}

The suggestion that M87 is the dominant source requires that our Galaxy
has a powerful magnetic wind:

\begin{itemize}
\item{}  implies Galactic wind with large size - 1/2 to 1 Mpc

\item{}  think of all galaxies with similar cosmic ray power, as measured
through star formation rate, and such through their far infrared
luminosity

\item{}  all similar galaxies should show such a wind bubble

\item{}  relatively strong magnetic field

\item{}  mass flow large, energy flow large, angular momentum flow small,
with a wind shell (observable in Lyman alpha absorption)

\item{}  the most important test of such a picture is to avoid the bottle
effect, thereby most high energy particles are screened out just as all
particles below a few hundred MeV are screened out from the Earth
environment by Solar modulation.  The strength and spatial distribution
of the $k^{-2}$ spectrum of magnetic irregularities have to allow this
condition - otherwise this picture fails completely

\end{itemize}

And, a fortiori, all galaxies with a star formation rate similar to ours
or even more, should have such a wind, building large cavities around
them.  If true, the sheets of the galaxy distribution should look like
slices of Swiss cheese in gas, with the ``holes" of order a Mpc across.
The next galaxy similar to ours in such a concept is probably M81, about 3
Mpc distant.

Recent simulations about propagation in our Galaxy also exist from
\cite{AES02,BGZ02}.

\section{Generalization:  Bottom-Up}

Nothing stated above is completely certain, we are at the beginning of a
beautiful era of deeper discovery, and so all the alternatives need to be
explored.

\begin{itemize}

\item{}  Distant radio galaxies with beam dump, such as 3C147, e.g.
\cite{Junor99}.  The production of neutrinos has generated a flurry of
contradictory papers recently,
\cite{LM2000,BaWa01,MPR01,KKSS02}

\item{}  Very high energy neutrinos: prediction (work by T. Weiler
\etal):  possible correlation with distant sources, characteristic
behaviour near horizon - could come from mountains near horizon.

\item{}  Magnetic monopoles, e.g., \cite{Wick02}?

\item{}  Prediction:  Correlation beyond GZK sphere

\item{}  other particles may be created in p-p collisions (e.g. work
by Farrar \etal)

\item{}  Pulsars???  Prediction:  Characteristic distribution in Galactic
plane, little correlation with supergalactic plane

\item{}  Many tests with those cosmic rays up to $3 \, 10^{18}$ eV

\end{itemize}

\section{Beautiful main alternative:  Top-down}

There is a whole world of models that suggests that the high energy
particles can be interpreted as decay products from even more energetic
or more massive particles, the top-down scenario, most intensely explored
recently by E.-J. Ahn, A. Olinto and G. Sigl.

\begin{itemize}

\item{}  Topological defects (work by Sigl \etal):  prediction:  rising
spectrum, large gamma ray background, \cite {Bhatta99,BS01,Si01a}

\item{}  If Quantum gravity scale near TeV:  then creation of small black
holes and branes possible in atmosphere, \cite{Ahn02a,Ahn02b,AFGS02}.

\item{}  Prediction:  Odd behaviour near horizon - \\ should come from
open space \\ in East (AUGER)

\item{}  different from ultra high energy neutrinos, \cite{AH01,HH02}

\item{}  if dark matter decay, then correlation with DM distribution,
e.g., \cite{BSS02a,BSS02b,BGG02}

\item{}  Radically New Physics? Lorentz Invariance Violation?
\cite{Protheroe2000,Ol01,Si01c,Si01d,Si01e,Ol02,Si02}

\end{itemize}

\section{Tasks}

All such predictions lead to specific tasks, which can be undertaken now
or in the near future.

\begin{itemize}
\item{}  Assume the various source classes, and test - include small
number graininess of source distribution

\item{}  Quiescent black holes, active black holes

\item{}  Test Galactic wind halo magnetic field

\item{}  Test supergalactic sheet magnetic field

\item{}  Test correlation with DM distribution

\item{}  Better understanding of the fluctuations

\item{}  Better understanding of fluorescence

\item{}  New experiments:  AUGER, EUSO, OWL

\item{}  New experimental techniques:  Radio emission:  LOFAR/SkyView,
\cite{FG02,HF02}; also infrared and acoustic detection methods may be
possible - after all most of the energy deposition of an air shower
almost certainly goes into heat, some into ionization, visible maybe
through radio absorption, and then some obviously into sound wave
generation

\item{}  new physics such as Lorentz Invariance violation.

\item{}  and many more...

\end{itemize}

\section{Conclusions}

\begin{itemize}

\item{}  Events need more statistics, at any energy

\item{}  Transition to Galactic cosmic rays

\item{}  Spectrum, arrival directions, chemical composition

\item{}  ?? New particles

\item{}  ?? Galactic magnetic field

\item{}  ?? Supergalactic magnetic field

\item{}  ?? Maximum energy observed

\item{}  ?? Detailed spectral features

\item{}  ?? Clustering of arrival directions

\item{}  ?? Properties of these clusters

\end{itemize}

Three main conclusions stand out from all the controversy surrounding the
current discussion; apart from just understanding the evolution of the
discussion, right back to the early years of last century:

\subsection{Experiment vs. experiment}
We absolutely need an experiment, that checks the different
techniques against each other; this will now be the AUGER experiment.
There are still fundamental questions, which make a direct comparison
between experiments rather difficult, such as ground array versus
fluorescence detector - AUGER will hopefully solve that.  However, we
desperately also need calibration from laboratory experiments, e.g. for
the fluorescence yield.  In order to understand experiments, theoreticians
need to develop the Monte-Carlo simulations, e.g. \cite{AM+02}, including
the large fluctuation statistics, and also the consistency tests.

\subsection{Theory vs. theory}
Before embarking on proposing a specific class of sources, all the
microphysics of such sources should be well developed, using the best
data from all other wavelengths.  Only when the observed spectrum has
been completely fitted, including temporal variability, with however
simple a physical model, can we proceed to ask what is the production of
very high energy particles, that may come all the way through to us.  The
cosmological evolution of stellar and non-stellar activity, the formation
and evolution of cosmic magnetic fields, all need to be included.
Especially cosmic magnetic fields need to be included, despite all the
uncertainties about their real strength and spatial distribution.

\subsection{Experiment vs. theory}
Selection effects and systematics can hamper any interpretation of data;
theoreticians should carefully listen to the experimentalists, but of
course with all due skepticism.  The best way is to always or at least
very often include an experimentalist as coauthor, or have an
experimentalist friend vet the paper, to make absolutely sure there is
mutual respect and common understanding of what an experiment can really
do, and what it maybe cannot do yet.  The ultimate solution may be to
intermingle theory and experiment in the education of our young students,
by sending them around, and including them very early on.

\subsection{Listen!}
We should really listen to Nature, what is the message that Nature is
giving us?  There is a message hidden in the various manifestations
of high energy particles, magnetic fields, interstellar and intergalactic
medium.  Nature speaks with a consistent voice.

What is the message that Nature is giving us?

\section{Acknowledgements}

P.L. Biermann would like to acknowledge the hospitality during his
prolonged stays, first at the University of Maryland, offered by his
colleague Eun-Suk Seo in the early spring of 2002, and then at the
University of Paris VII, in the late spring 2002, offered by his
colleagues Norma Sanchez and Hector de Vega.  Work with PLB is being
supported through the AUGER theory and membership grant 05
CU1ERA/3 through DESY/BMBF (Germany); further support for the work with
PLB comes from the DFG, DAAD, Humboldt Foundation (all Germany),
grant 2000/06695-0 from FAPESP (Brasil) through G. Medina-Tanco, KOSEF
(Korea), and ARC (Australia).  Important input to the discussions in this
paper has come from E.-J. Ahn, H. Bl{\"u}mer, M. Chirvasa, A.C. Donea, R.
Engel, T. En{\ss}lin, H. Falcke, C. Galea, N. Ikhsanov, K.-H. Kampert, H.
Kang, M. Kaufman, P.P. Kronberg, N. Langer, H. Lee, E. Loh, S. Markoff,
A. Meli, S. Moiseenko, F. Munyaneza, B. Nath, A. Popescu, R.J. Protheroe,
G. Pugliese, W. Rhode, G. Romero, E. Roulet, D. Ryu, N. Sanchez, E.-S.
Seo, G. Sigl, T. Stanev, S. Ter-Antonyan, M. Teshima, P. Tinyakov, A.
Vasile, H. de Vega, Y. Wang, A. Watson, T. Weiler, St. Westerhoff, and
Ch. Zier.

\end{document}